\documentclass[10pt, paper=a4, UKenglish]{article}
\usepackage{graphicx}
\usepackage{amsmath}
\usepackage{amsfonts}
\usepackage{caption}
\usepackage{algorithm}
\usepackage{booktabs}
\usepackage{isotope}
\usepackage{multirow}
\usepackage[noend]{algpseudocode}

\def\Title#1{\begin{center} {\Large #1 } \end{center}}
\def\Author#1{\begin{center}{ \sc #1} \end{center}}
\def\Address#1{\begin{center}{ \it #1} \end{center}}

\newcommand\pubblock{\rightline{\begin{tabular}{l} Proceedings of the CTD/WIT 2019\\ \pubnumber\\
         \pubdate  \end{tabular}}}

\newenvironment{Abstract}{\begin{quotation} \begin{center} 
             \large ABSTRACT \end{center}\bigskip 
      \begin{center}\begin{large}}{\end{large}\end{center} \end{quotation}}

\newenvironment{Presented}{\begin{quotation} \begin{center} 
             PRESENTED AT\end{center}\bigskip 
      \begin{center}\begin{large}}{\end{large}\end{center} \end{quotation}}

\def\Acknowledgements{\bigskip  \bigskip \begin{center} \begin{large}
      \bf ACKNOWLEDGEMENTS \end{large}\end{center}}

\graphicspath{{img/}}

\def\beq{\begin{equation}}
\def\eeq#1{\label{#1}\end{equation}}
\def\eeqn{\end{equation}}

\def\beqa{\begin{eqnarray}}
\def\eeqa#1{\label{#1}\end{eqnarray}}
\def\eeqan{\end{eqnarray}}

\let\bar=\overbar

\def\Dslash{\not{\hbox{\kern-4pt $D$}}}
\def\dslash{\not{\hbox{\kern-2pt $\del$}}}

\def\msb{{\bar{\ssstyle M \kern -1pt S}}}

\usepackage{color}
\usepackage{lineno}
\usepackage{subfig}
\usepackage{hyperref}

\newcommand\pubnumber{PROC-CTD19-012}

\newcommand\pubdate{\today}

\def\affiliation{
On behalf of\\\vspace{0.2em}
\textsuperscript{1}Institute of Experimental and Applied Physics\\
Czech Technical University, Prague, Czech Republic\\\vspace{0.2em}
and\\\vspace{0.2em}
\textsuperscript{2}Faculty of Electrical Engineering,\\
University of West Bohemia, Pilsen, Czech Republic}

\newcommand{\conference}{Connecting the Dots and Workshop on Intelligent Trackers (CTD/WIT 2019)\\
Instituto de F\'isica Corpuscular (IFIC), Valencia, Spain\\ 
April 2-5, 2019}

\usepackage{fancyhdr}
\definecolor{mygrey}{RGB}{105,105,105}
\begin{document}


\large
\begin{titlepage}
\pubblock

\vfill
\Title{Randomized Computer Vision Approaches for Pattern Recognition\\in Timepix and Timepix3 Detectors}
\vfill

\Author{Petr M\'anek\textsuperscript{1}, Benedikt Bergmann\textsuperscript{1}, Petr Burian\textsuperscript{1,2},\\Luk\'a\v{s} Meduna\textsuperscript{1}, Stanislav Posp\'i\v{s}il\textsuperscript{1}, Michal Suk\textsuperscript{1}}
\Address{\affiliation}
\vfill

\begin{Abstract}
Timepix and Timepix3 are hybrid pixel detectors ($256\times 256$ pixels), capable of tracking ionizing particles as isolated clusters of pixels. To efficiently analyze such clusters at potentially high rates, we introduce multiple randomized pattern recognition algorithms inspired by computer vision. Offering desirable probabilistic bounds on accuracy and complexity, the presented methods are well-suited for use in real-time applications, and some may even be modified to tackle trans-dimensional problems. In Timepix detectors, which do not support data-driven acquisition, they have been shown to correctly separate clusters of overlapping tracks. In Timepix3 detectors, simultaneous acquisition of Time-of-Arrival (ToA) and Time-over-Threshold (ToT) pixel data enables reconstruction of the depth, transitioning from 2D to 3D point clouds. The presented algorithms have been tested on simulated inputs, test beam data from the Heidelberg Ion therapy Center and the Super Proton Synchrotron and were applied to data acquired in the MoEDAL and ATLAS experiments at CERN.
\end{Abstract}

\vfill

\begin{Presented}
\conference
\end{Presented}
\vfill
\end{titlepage}
\def\thefootnote{\fnsymbol{footnote}}
\setcounter{footnote}{0}
%

\normalsize 


\section{Introduction}%
\label{intro}%
Timepix~\cite{llopart2007timepix} are hybrid pixel detectors developed by the Medipix collaboration, CERN~\footnote{\url{https://medipix.web.cern.ch/}}. They are comprised of a semiconductive sensor layer, uniformly divided into a square matrix of $256\times 256$ pixels with $55\ \mathrm{\mu m}$ pitch. Timepix3~\cite{poikela2014timepix3} are successor models of Timepix. Keeping the same pixel layout, they offer superior resolution of time and energy. While Timepix are operated in frame-based mode, producing 2D images integrated over the data acquisition period, Timepix3 also support event-based readout, wherein the detector asynchronously reports hits when they are observed. Detectors of the Timepix family have variety of applications in
life sciences~\cite{jakuubek2009semiconductor} and hadron therapy~\cite{hartmann2012towards}. With great promise for space radiation monitoring~\cite{filgas2018space}, Timepix-based systems were mounted at ISS~\cite{turecek2011small,stoffle2015timepix} and on board the SATRAM orbital platform~\cite{bergmann2016measurement,gohl2019study}. To help characterize mixed radiation fields, Timepix and Timepix3 detector networks were installed at the ATLAS~\cite{bergmann2016atlas} and MoEDAL~\cite{pinfold2017moedal}, experiments at the Large Hadron Collider. In all of the listed applications, researchers relied on automated methods to efficiently analyze large quantities of measured data.

Prior to analysis, outputs of the data acquisition process are usually aggregated. While Timepix3 data only require reordering, Timepix frames are subdivided into characteristic patterns called \textit{clusters}, each corresponding to a hit. Since often a multitude of clusters of various shapes are present in a single frame, their segmentation and classification poses a non-trivial task, which is crucial to any subsequent work. In the past, clusters were processed morphologically~\cite{holy2008pattern}, relying on connectivity-checking algorithms (e.g.~Flood Fill) to isolate clusters as spatially localized groups of active pixels, and producing classifications by thresholding on various morphological properties (e.g.~linearity, convex hull, etc.). This however imposes additional constraint on the sparsity of processed frames, expecting that only pixels corresponding to common incidence events may be assumed to be spatially adjacent.

The sparsity requirement is often easily satisfied by tuning acquisition duration with respect to observed frame occupancy. This procedure may however pose a challenge in unpredictable and high-flux environments (e.g.~in space). In addition, for short frame durations data acquisition is dominated by the constant dead time of 11~ms, during which pixel data are read out and the detector is not sensitive. Having been originally developed for predecessor detectors capable only of binary discrimination in each pixel, morphological segmentation may also be viewed as inherently suboptimal: with quantized intensity information available in Time-over-Threshold (ToT) mode, Timepix frames are reduced to binary images, effectively thresholding intensity values and resulting in precision loss. With that in mind, the main objective of this work is to introduce novel analysis methods suitable for use as alternatives to conventional morphological approaches.

\section{Methods}%
This contribution proposes new analysis methods for Timepix and Timepix3 data, often inspired by recent developments in the field of computer vision. Unlike their morphological counterparts, these approaches may be considered advantageous since they do not require thresholding of intensities at input, and can thus potentially achieve better accuracies~\cite{soukup2012dynamics}. In addition, such algorithms usually posess desirable complexity properties or offer adjustable trade-off between complexity and accuracy (useful e.g. for real-time processing of camera frames). The presented methods are based on robust algorithms which have been shown to reliably diminish adverse effects of input noise, often in relatively large quantities.

In order to apply such methods to Timepix data, frames are effectively treated as images, where pixel intensity refers to the amount of deposited energy (in keV) calibrated from quantized ToT values~\cite{jakubek2011precise}. Cluster segmentation can then be viewed as an object detection task. With that in mind, this work focuses primarily on a specific class of high energy transfer events constituted by heavy ions.

\begin{figure}[!htb]
  \centering
  \includegraphics[width=0.48\linewidth,trim={0.5em 0.5em 1.2em 1.7em},clip]{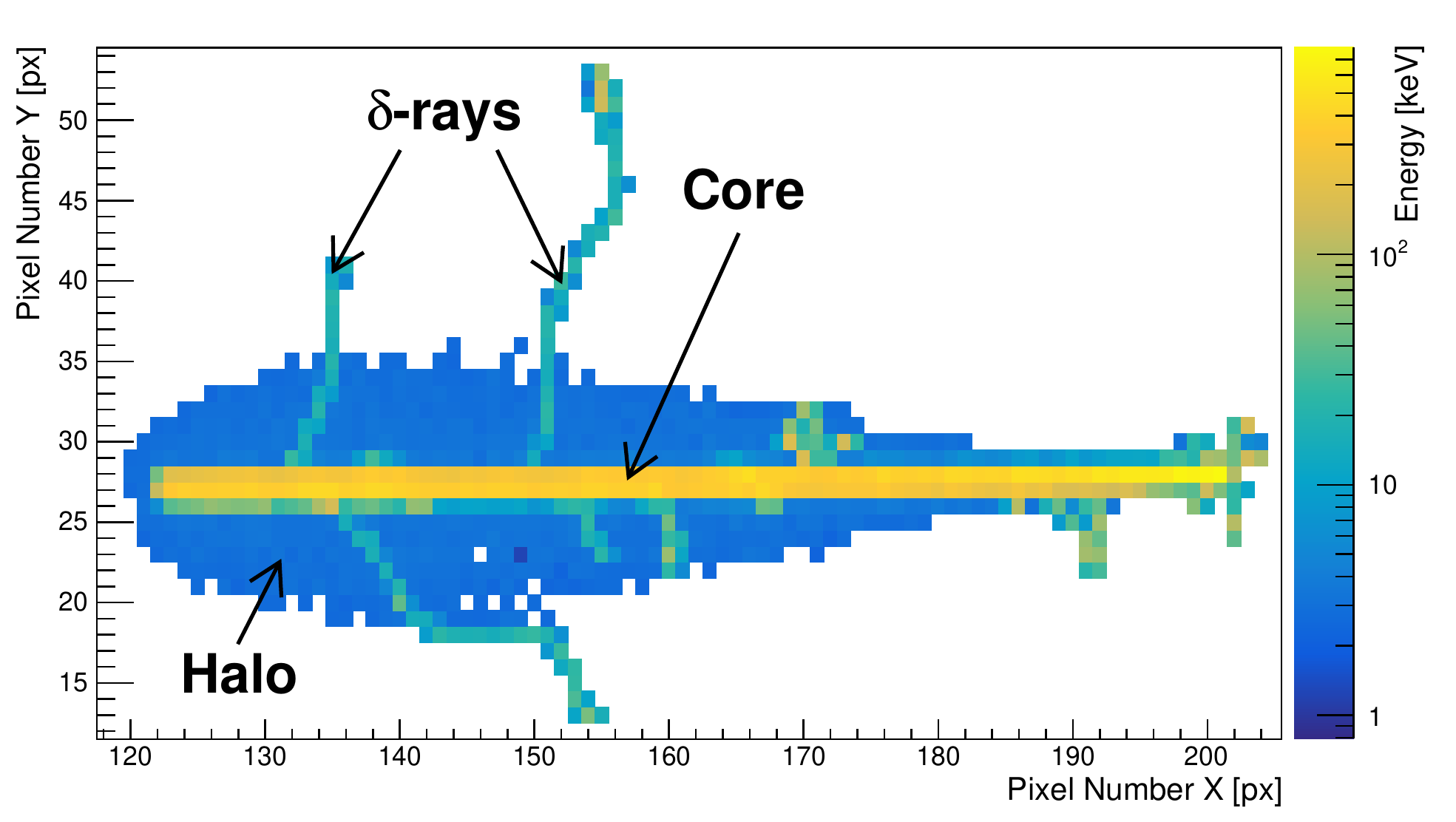}
  \caption{Cluster corresponding to Ar ion at 75~GeV/c with characteristic regions annotated.}
  \label{fig:annotated-cluster}
\end{figure}

Timepix clusters produced by heavy ions (shown in Figure~\ref{fig:annotated-cluster}) are usually labeled as heavy tracks or heavy blobs by morphological classification. Conventionally, three regions of interest are distinguished: (1) a high-energy linear core carrying trajectory information, (2) a surrounding low-energy halo and (3) a multitude of $\delta$-rays probabilistically emitted from the core outwards.

\subsection{Randomized Trajectory Fitting}%
The first presented method efficiently estimates the spatial parameter model of the particle trajectory by means of randomized fitting. It is based on the well-known RANSAC algorithm~\cite{fischler1981random}. RANSAC has been shown to possess robust behavior and advantageous complexity properties, in some cases even reaching processing rates suitable for real-time computer vision applications. Since each iteration of the algorithm represents an independent instance of random sampling, the total number of iterations is related to the expected probability of a correct sampling. Given a desired confidence threshold, this conveniently allows to assess fit quality at runtime, enabling early termination once a sufficient fit quality is reached.

Due to the large momentum of heavy ions, a linear trajectory can be assumed (i.e. a line segment). Fit parameters are two planar points $\vec{a},\vec{b}\in\mathbb{R}^2$ representing the entry point and the exit point of the particle trajectory in the sensor layer, respectively. Within a set of hit pixels, a sampled subset of size 2 is required to fully determine an instance of such model. To formally define the fitting optimization task, energy maximizer formulation is used. The goal is to maximize the utility:
\begin{align}
  U_{w,\Phi}(\mathcal{M}\mid F)
  =
  \int_{s=0}^{1}
  \int_{t=-t_\mathrm{max}}^{t_\mathrm{max}}
  w(|t|)
  \Phi (f_F((1-s)\vec{a}+s\vec{b}+t\vec{n}))
  dt ds.
  \label{fml:ransac-utility}
\end{align}
Here, $t_\mathrm{max}$ is a sufficiently large distance (e.g. cluster diameter), $\vec{n}$ is a normalized 2D vector orthogonal to $\vec{b}-\vec{a}$, $f_F:\mathbb{R}^2\rightarrow \mathbb{R}_0^+$ is a Timepix frame image function, $w:\mathbb{R}_0^+\rightarrow [0,1]$ is a distance weight function and $\Phi :\mathbb{R}_0^+\rightarrow\mathbb{R}$ is an energy kernel function.

For the purpose of comparison, three RANSAC-inspired algorithms were developed. Their general scheme is given by Algorithm~\ref{alg:ransac-single}. The proposed algorithms are labeled as follows:
\begin{description}
  \item[RANSAC]
  Conservative implementation, wherein sampled pixels are directly used for model fitting. 
  
  \item[LO-RANSAC]
  Extension of RANSAC which attempts to improve new models using greedy local optimizer.

  \item[SA-RANSAC]
  Extension of LO-RANSAC, wherein simulated annealing is used to accelerate computationally intensive local optimization process~\cite{SimulatedAnnealing}.
\end{description}

\begin{algorithm}
  \caption{RANSAC, LO-RANSAC and SA-RANSAC}~\label{alg:ransac-single}
  \begin{algorithmic}[1]
          \State $r_\text{max}\gets \text{maximum number of iterations}$
          \State $U^*\gets -\infty$, $r\gets 1$
          \Repeat
          \State Select pixel coordinates $\vec{a}^r, \vec{b}^r$ at random.
          \State Fit model $\mathcal{M}^r$ from $\vec{a}^r, \vec{b}^r$.
          \State Evaluate the utility of the model $U^r\gets U_{w,\Phi}(\mathcal{M}^r\mid F)$.
          \If{$U^r>U^*$}
                  \State Improve $U^*\gets U^r$, $\mathcal{M}^*\gets\mathcal{M}^r$ (with the help of no / greedy / SA optimizer).
          \EndIf
          \State $r\gets r+1$
          \Until{$r\geq r_\text{max}$} 
  \end{algorithmic}
\end{algorithm}

\subsection{Decoupled Cluster Segmentation}%
The second presented method aims to improve the segmentation accuracy in saturated frames by exploiting intensity information and linear appearance of overlapping tracks. To achieve this purpose, multiple algorithms are employed in a decoupled scheme to propose and gradually refine segmentation into its final form. For initialization, a conventional morphological method is used. Clusters are then independently analyzed by Hough Transform~\cite{duda1971use} to detetermine the likelihood of insufficient partitioning, and to produce secondary segmentation if necessary.

Hough Transform is a computer vision method often used for recognition of complex parametric shapes and patterns (e.g. lines, circles, etc.). Its major advantages include efficient tractability in 2D and robustness towards relatively large amount of input noise. In the presented algorithm, a version capable of detecting lines is utilized with the motivation to recognize integrated track overlaps and piecewise linear prongs. To avoid problems caused by possibly unbounded values of conventional linear parameterizations, the transform uses \textit{Hesse normal form} which parameterizes lines as:
\begin{align}
        r = x_1\cos\theta + x_2\sin\theta
        \qquad
        \text{where }
        \theta\in [0,\pi],
        r\in[-r_\text{max},r_\text{max}]
\end{align}

In a modified algorithm (given by Algorithm~\ref{alg:hough-ransac-lo}), after the accumulator is populated with votes from each pixel of the analyzed cluster, global maxima addressed by $\theta$ and $r$ are identified in the accumulator. Each such maximum represents a linear fragment of the cluster, and is removed along with its local neighborhood both from the cluster and the accumulator. This way, duplicate detections are prevented without the necessity for subsequent non-maxima suppression. This process continues sequentially until the value of the maximum falls below an adjustable threshold.

\begin{algorithm}
  \caption{Segmentation Refinement by Hough Transform}~\label{alg:hough-ransac-lo}
  \begin{algorithmic}[1]
          \State $A_\text{min}\gets \text{accumulator termination threshold}$, $s_\text{max}\gets \text{maximum number of iterations}$
          \State $A\gets \text{empty accumulator}$, $s\gets 1$
          \For{$(x,e)\in F$}\Comment{Each pixel votes for all lines consistent with it.}
          \For{$(\theta,r)\in\mathcal{L}(x)$}
          \State Cast vote $A(\theta,r)\gets A(\theta,r)+\Phi(e)$.
          \EndFor
          \EndFor
          \Repeat\Comment{Most likely maxima are sequentially removed.}
          \State $(\theta^s,r^s)\gets\text{argmax}_{(\theta,r)} A(\theta,r)$
          \State $A^s\gets A(\theta^s,r^s)$
          \State $C_s\gets\{(x,e)\in F\mid x\text{ consistent with }(\theta^s,r^s)\}$
          \For{$(x,e)\in C_s$}\Comment{Votes for the removed line are subtracted.}
          \For{$(\theta,r)\in\mathcal{L}(x)$}
          \State Suppress inlier vote $A(\theta,r)\gets A(\theta,r)-\Phi(e)$.
          \EndFor
          \EndFor
          \State Report subset $C_s$.
          \State $s\gets s+1$
          \Until{$A^s\leq A_\text{min}$ or $s\geq s_\text{max}$}
  \end{algorithmic}
\end{algorithm}

\subsection{Particle Identification}%
The goal of the third presented method is to achieve particle identification (PID) by sampling and aggregating intensities from frames. This is motivated by the theory of energy deposition along the particle trajectory within a Timepix sensor. In general, this quantity was described by the Bethe-Bloch equation~\cite{ReviewOfPhysics} as sensitive to the nuclear charge of the projectile and its velocity. To achieve classification capability, this relationship is exploited by a simple machine learning algorithm.

The stopping power is sampled from clusters across several uniform intervals along the particle trajectory (as fitted by e.g. RANSAC). Sampled energies with normalized ordering constitute a feature model, which is used to train a standard $k$-NN classifier~\cite{KNNold}. With a sufficient number of examples, the classifier learns stopping powers of various particle classes, gaining the ability to label previously unobserved clusters by cross-referencing their features with internal representation of training data.

\section{Results}%
For evaluation, both experimental and artificial datasets were utilized. Even though testing with experimental data is generally preferable as it provides valuable insight into behavior under realistic conditions, inherent lack of ground truth information complicates the quantification of spatial and angular accuracies. The generation of entirely artificial datasets (e.g. by means of a physics simulation) on the other hand fails to reproduce phenomena relevant to evaluated algorithms. For these reasons, a middle ground approach was selected, wherein experimental data were curated, manually annotated with approximate ground truth information and randomly combined, producing large artificial datasets of realistic Timepix frames.

The experimental data used was measured at the Heidelberg Ion Therapy Center\footnote{Heidelberger Ionenstrahl-Therapiezentrum (HIT), Im Neuenheimer Feld 450, 69120 Heidelberg, \url{https://www.klinikum.uni-heidelberg.de/index.php?id=113005&L=1}} and Super Proton Synchrotron, CERN. The setup included a Timepix detector equipped with a 1~mm thick Si sensor layer. The bias voltage was set to 500~V and the acquisition time was set to 0.1~s. During the experiment, charged particles hit the detector at various angles. Each dataset hence corresponds to a specific beam configuration (particle species and momentum) and incidence angle. In the datasets, the following particle species are represented: $p$, He, C, O, Ar. To increase quality, measured datasets were first automatically processed to exclude corrupted, oversaturated and noisy frames. Out of several thousands remaining frames, the total of 937 ground truth tracks were manually prepared.

\subsection{Randomized Trajectory Fitting}%
The first experiment examines the behavior of the presented RANSAC-inspired algorithms in a very basic setting comprised of frames containing only one track produced by a single particle. Although such a scenario is not likely to be often observed in reality, it is desirable for the purposes of evaluation. As the lack of overlaps renders segmentation unnecessary, it provides a valuable insight into the performance of proposed trajectory fitting methods.

\begin{figure}[!htb]
  \centering
  \includegraphics[width=0.3\linewidth,trim={0 51em 38em 2em},clip]{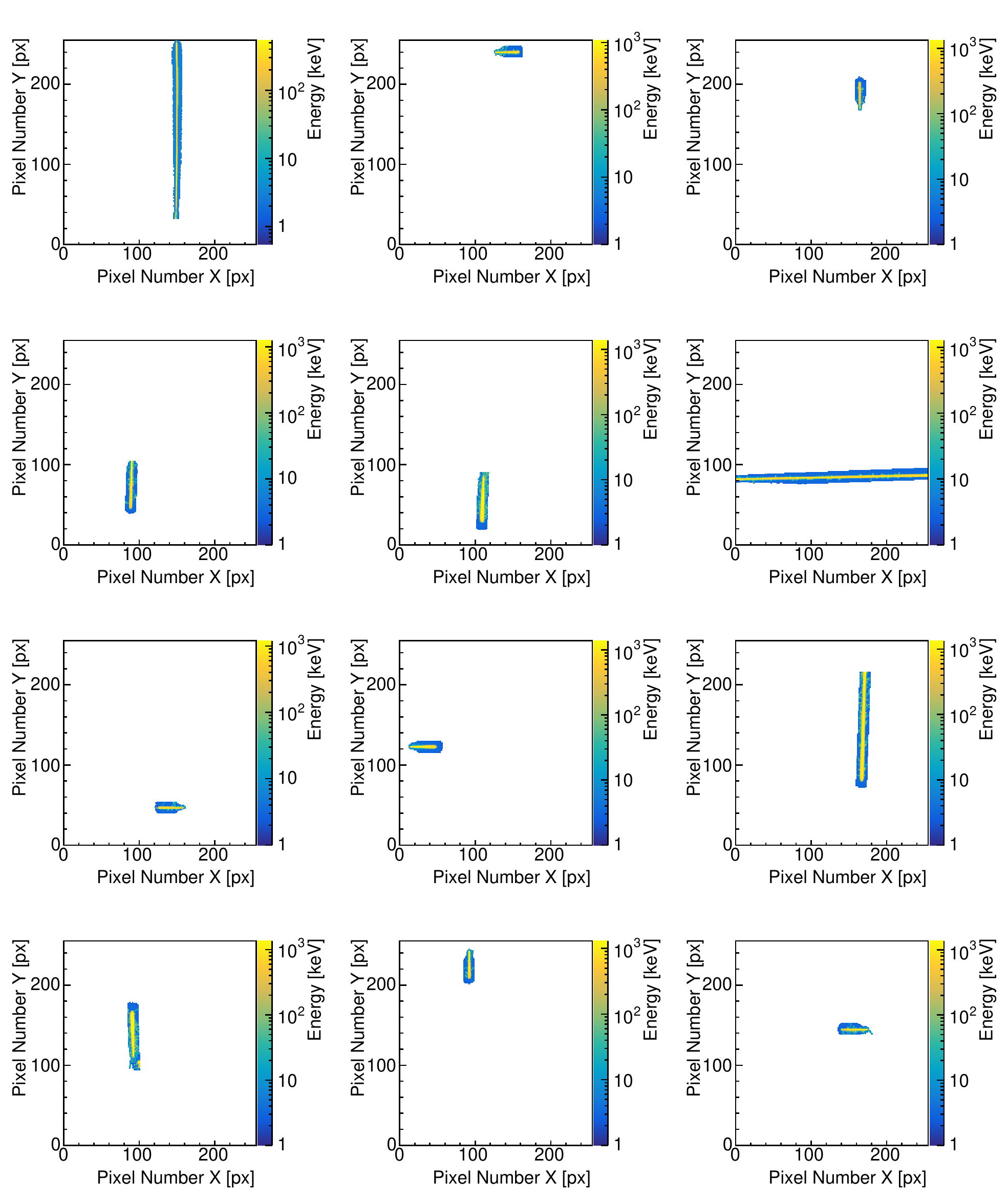}
  \quad
  \includegraphics[width=0.3\linewidth,trim={0 17em 38em 36em},clip]{sp_frames}
  \quad
  \includegraphics[width=0.3\linewidth,trim={0 0 38em 53em},clip]{sp_frames}
  \caption{Selected artificial frames used for trajectory fitting evaluation.}
  \label{fig:single-particle-data}
\end{figure}

For the experiment, 1,000 artificial frames were generated from the ground truth dataset. This quantity is considered to be a good compromise between tractability and accuracy. In each frame, a track was chosen uniformly at random, and placed at a random non-cropping location within the pixel matrix. To suppress directional bias, track orientation was randomized as well. However, in order to avoid intensity interpolation, the rotation angle was quantized to 90 degrees. Several examples of the generated frames are shown in Figure~\ref{fig:single-particle-data}.

In evaluation of the presented algorithms, two standard metrics are tracked -- the failure rate $FR$ and the error $E(v)$ of arbitrary parameter $v$. These metrics are conservatively defined as follows:
\begin{align}
  FR=\frac{FP + FN}{TP + TN + FP + FN}
  \qquad\qquad
  E(v)=
  \sqrt{
  \frac{1}{n}
  \sum_{i=1}^n
  (v_i-\hat{v}_i)^2
  }
\end{align}
Here, the values $FP$, $FN$, $TP$, $TN$ refer to the numbers of false positive, false negative, true positive and true negative detections, respectively. Moreover, for $i=1,2,\dots n$ the sequences $\hat{v}_i$ and $v_i$ represent fitted parameters and their corresponding ground truth values, respectively. Since there is exactly one track in each frame, zero detections are regarded as a false negative, one detections as a true positive, and any further detections fits as false positives. To suppress random effects, each frame has been repeatedly processed 3~times. The results shown in Table~\ref{tbl:single-particle-detection} are mean values over all runs. The values of used configuration parameters are shown in Table~\ref{tbl:overlap-detection}.

\begin{table}[!htb]
  \centering
  \small
  \begin{tabular}{l|r|rrrr|r}
  \toprule
  Algorithm & $\left<FR\right>$ & $\left<E(a_1)\right>$ & $\left<E(a_2)\right>$ & $\left<E(\varphi)\right>$ & $\left<E(\theta)\right>$ & $\left<t\right>$ [ms]\\
  \midrule
  RANSAC & 0.022 & 3.227 & 2.821 & 3.047 & 10.412 & 1,012.7 \\
  LO-RANSAC & 0.021 & 2.898 & 2.873 & 2.886 & 11.863 & 5,538.6 \\
  SA-RANSAC & 0.021 & 2.721 & 2.908 & 2.856 & 12.582 & 1,762.5 \\
  \bottomrule
  \end{tabular}
  \caption{Results of trajectory fitting evaluation.}
  \label{tbl:single-particle-detection}
\end{table}

\subsection{Decoupled Cluster Segmentation}%
\begin{figure}[!htb]
  \centering
  \includegraphics[width=0.3\linewidth,trim={0 51em 38em 2em},clip]{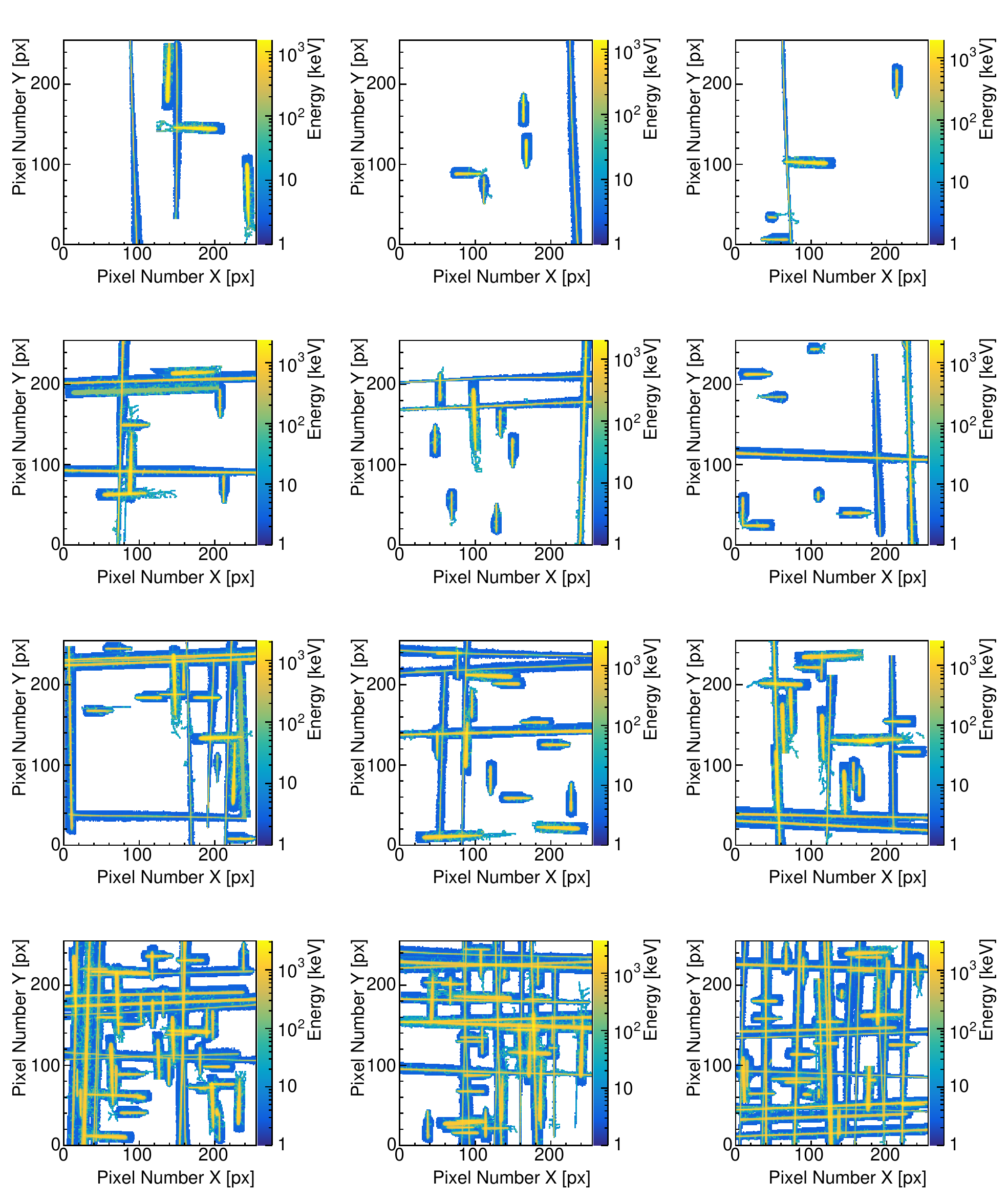}
  \quad
  \includegraphics[width=0.3\linewidth,trim={0 17em 38em 36em},clip]{mp_frames}
  \quad
  \includegraphics[width=0.3\linewidth,trim={0 0 38em 53em},clip]{mp_frames}
  \caption{Selected artificial frames used for overlap segmentation evaluation with varying track counts. From left to right, $n=5,20,50$ respectively.}
  \label{fig:multiple-particle-data}
\end{figure}

In the second experiment, the enhanced overlap segmentation capability of the presented decoupled segmentation algorithm was evaluated. In a familiar scheme, artificial datasets were randomly generated from manually annotated measured data. In this instance, however, the artificial frames contain varying numbers of overlapping tracks $n_\text{trk}\in\{5,10,20,50\}$ (shown in Figure~\ref{fig:multiple-particle-data}). These can be viewed as simulations taken with increasing acquisition time.

In evaluation of every frame, a matching between ground truth detections and produced detections was constructed, wherein only detections with sufficiently small distance were matched. While successfully matched pairs were regarded as true positives, unmatched ground truth detections were treated as false negatives and unmatched detections provided were considered to be false positives.

To provide realistic insight into practical application of all presented methods, detections produced by Hough segmentation were additionally processed by RANSAC methods. This way, any systematic fitting bias introduced by the Hough method is amplified in contrast to the results of the previous experiment. Randomized evaluation was performed in the same scheme (1,000 frames, 3 runs each). The results as well as values of used configuration parameters are shown in Table~\ref{tbl:overlap-detection}.

\begin{table}[!htb]
  \centering
  \subfloat{\small\begin{tabular}{rr|r|rrrr|r}
    \toprule
    {} & $n_\text{trk}$ & $\left<FR\right>$ & $\left<E(a_1)\right>$ & $\left<E(a_2)\right>$ & $\left<E(\varphi)\right>$ & $\left<E(\theta)\right>$ & $\left<t\right>$ [ms]\\
    \midrule
    \parbox[t]{1mm}{\multirow{4}{*}{\rotatebox[origin=c]{90}{\tiny RANSAC}}} & 5 & 0.107 & 4.344 & 4.266 & 5.244 & 10.085 & 4,492.2  \\
    {} & 10 & 0.253 & 6.413 & 6.384 & 9.030 & 10.826 & 9,123.3  \\
    {} & 20 & 0.544 & 10.264 & 9.776 & 12.012 & 11.786 & 20,626.7  \\
    {} & 50 & 0.772 & 14.852 & 14.572 & 14.655 & 10.226 & 38,161.3  \\
    \midrule
    \parbox[t]{1mm}{\multirow{4}{*}{\rotatebox[origin=c]{90}{\tiny LO-RANSAC}}} & 5 & 0.104 & 4.194 & 3.953 & 5.169 & 11.709 & 24,105.5  \\
    {} & 10 & 0.242 & 6.096 & 6.219 & 8.549 & 12.250 & 57,995.2  \\
    {} & 20 & 0.532 & 9.777 & 9.311 & 11.620 & 12.682 & 125,795.0  \\
    {} & 50 & 0.770 & 14.034 & 13.888 & 13.960 & 10.167 & 294,498.0  \\
    \midrule
    \parbox[t]{1mm}{\multirow{4}{*}{\rotatebox[origin=c]{90}{\tiny SA-RANSAC}}} & 5 & 0.105 & 4.026 & 3.938 & 5.115 & 12.205 & 20,882.6  \\
    {} & 10 & 0.238 & 6.088 & 6.064 & 7.833 & 12.539 & 19,708.4  \\
    {} & 20 & 0.528 & 9.768 & 9.163 & 10.912 & 12.578 & 39,827.8  \\
    {} & 50 & 0.769 & 14.324 & 14.111 & 12.535 & 9.892 & 83,365.4  \\
    \bottomrule
    \end{tabular}}%
  \quad
  \subfloat{\small\begin{tabular}{ll}
    \toprule
    \multicolumn{2}{c}{\footnotesize RANSAC / LO-RANSAC / SA-RANSAC}\\
    \midrule
    Parameter & Value\\
    \midrule
    Energy Kernel $\Phi$ & ReLU\\
    Distance Weight $w$ & Gaussian\\
    Sample Count & 2,000\\
    Connectivity Thl. $e_0$ & $10^{-3}$ keV\\
    Neighborhood Size & 8\\
    L.O. Iterations & 0 / $10^2$ / $10^3$\\
    L.O. Window $\Delta$ & $\{0,\pm 0.2\}$ \\
    Annealing Schedule & 0 / 0 / linear\\
    \midrule
    H.T. Inlier Distance & 7 pixels\\
    H.T. Minimum Thl. & 5,000\\
    H.T. Iterations & 100\\
    \bottomrule
    \end{tabular}}
  \caption{Results of overlap detection evaluation (left) and the configuration parameters used (right).}
  \label{tbl:overlap-detection}
\end{table}

\subsection{Particle Identification}%
To investigate the viability of the presented feature model for particle identification, the Heidelberg datasets were used once again. Unlike in previous sections, here ground truth information consists only of the particle species, a known property of the particle beam, eliminating any necessity for manual data labeling.

\begin{figure}[!htb]
  \centering
  \subfloat{\includegraphics[height=5.7cm,trim={0 0 100px 25px},clip]{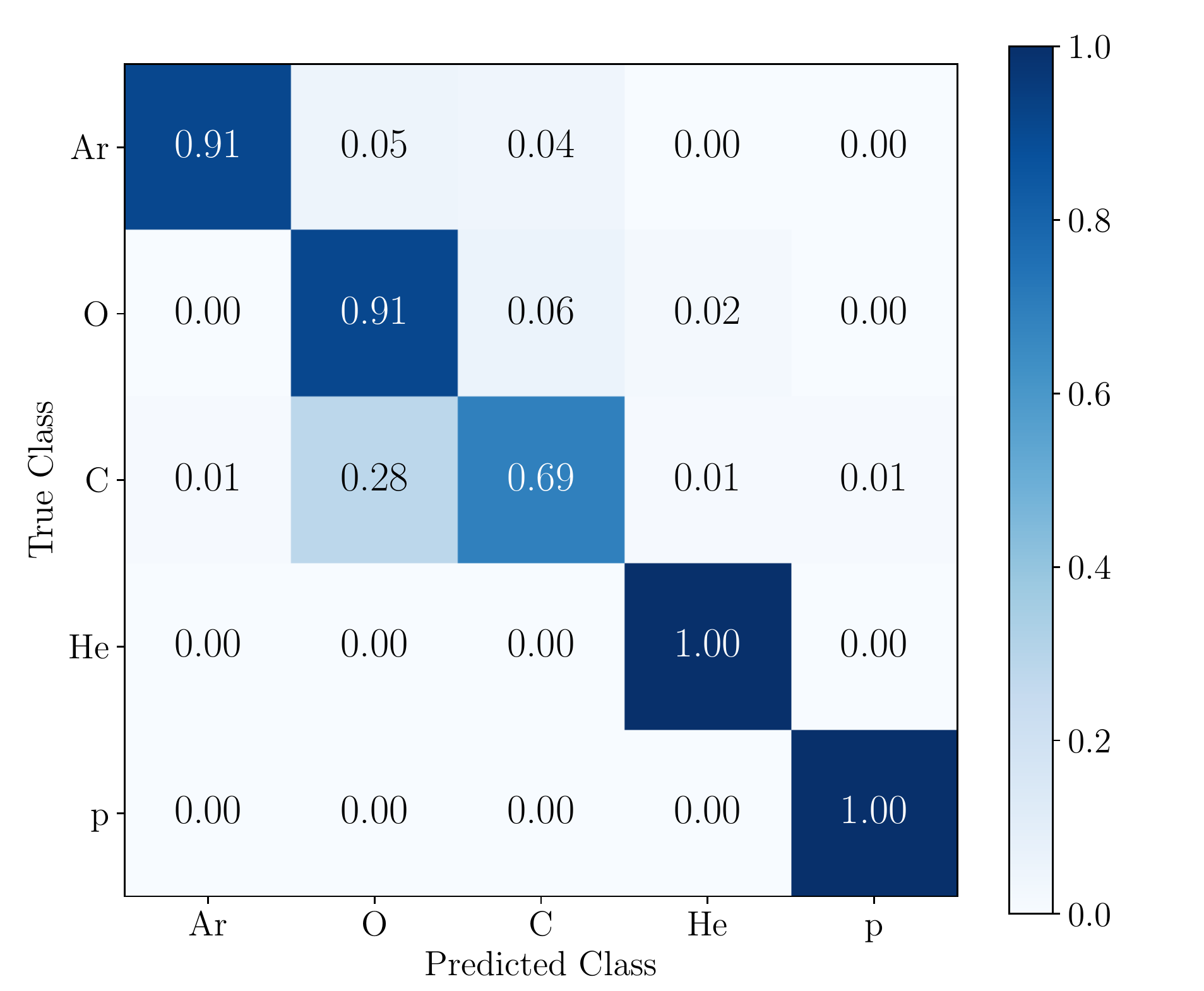}}
  \quad
  \subfloat{\includegraphics[height=5.7cm,trim={0 0 130px 33px},clip]{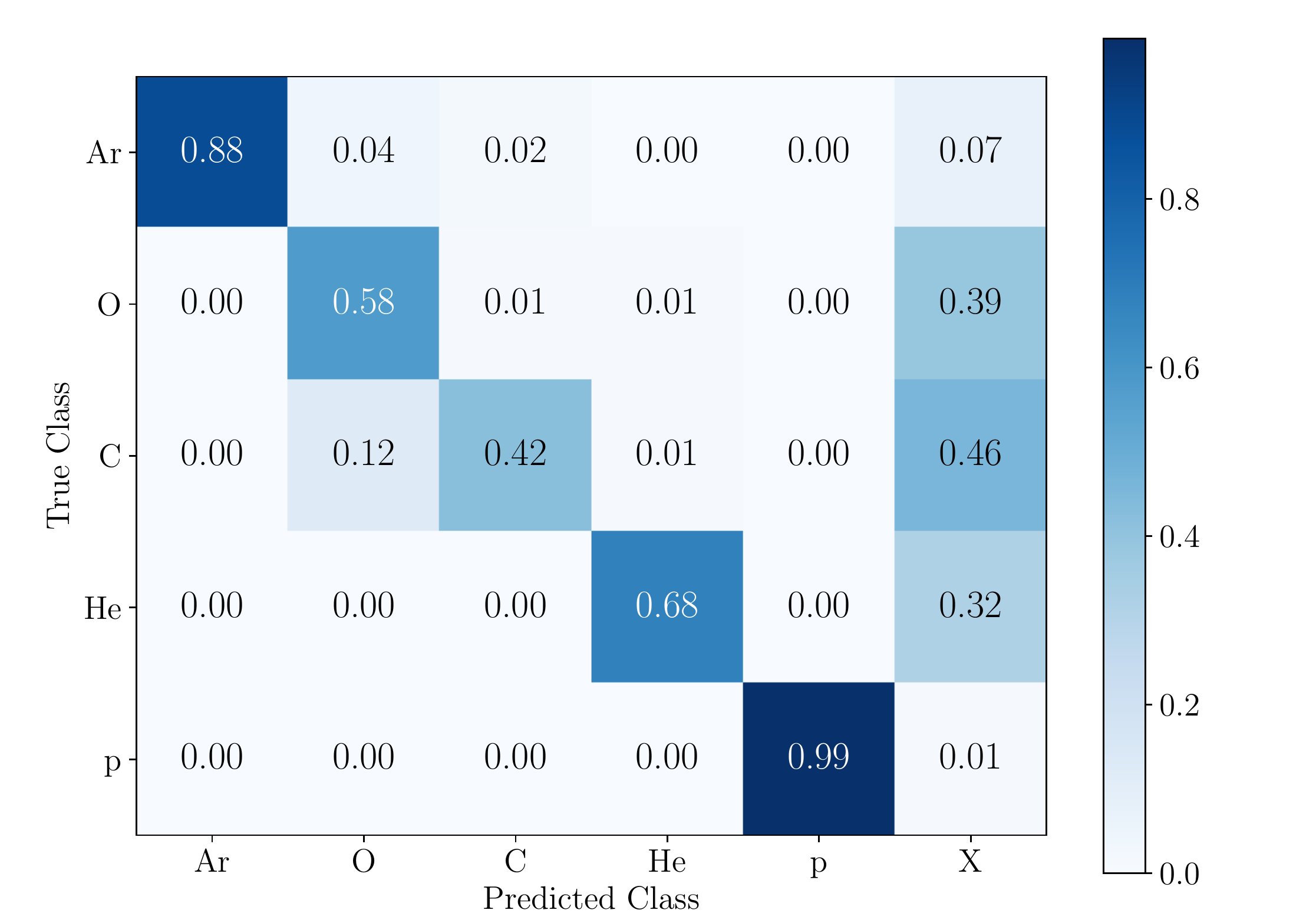}}
  \caption{Confusion matrices of the $k$-NN classifier ($k=7$, $n_\text{int}=16$) without rejection (left, accuracy~=~0.893) and with rejection (right, $c=0.9$, accuracy~=~0.935, RR~=~0.249).}
  \label{fig:confusion-matrices}
\end{figure}

For evaluation, a stratified $K$-fold cross-validation scheme was adopted. The experimental data was randomly divided into 5 equally-sized folds, each containing approximately equal representation of all classes. In multiple runs, four folds were interpreted as the training set, whereas the fifth served as the testing set, producing a single accuracy score per run. Finally, the overall accuracy score was calculated as the mean of scores from all runs. This way, the effects of the hyperparamter $n_\text{int}$ determining the number of sampling intervals were investigated for values $n_\text{int}\in\{1,16,32,64,128\}$. The calculated accuracies are shown in Table~\ref{tbl:particle-identification} and Figure~\ref{fig:confusion-matrices}.

\begin{table}[!htb]
  \centering
  \subfloat{\small\begin{tabular}{r|rr|rr}
  \toprule
  $n_\text{int}$ & $\left<t_\text{train}\right>$ [ms] & $\left<t_\text{test}\right>$ [ms] & $\left<\text{accuracy}\right>$ \\
  \midrule
  1 & 0.906 & 1.054 & 0.851 \\
  16 & 0.645 & 1.917 & 0.893 \\
  32 & 1.023 & 5.315 & 0.892 \\
  64 & 2.051 & 6.065 & 0.885 \\
  128 & 1.901 & 11.225 & 0.887 \\
  \bottomrule
  \end{tabular}}
  \quad
  \subfloat{\small\begin{tabular}{r|rr|rr}
  \toprule
  $c$ & $\left<t_\text{train}\right>$ [ms] & $\left<t_\text{test}\right>$ [ms] & $\left<\text{accuracy}\right>$ & $\left<\text{RR}\right>$ \\
  \midrule
  0.50 & 1.208 & 2.481 & 0.895 & 0.006 \\
  0.75 & 0.731 & 2.325 & 0.922 & 0.111 \\
  0.90 & 0.620 & 1.612 & 0.935 & 0.249 \\
  0.95 & 0.721 & 2.398 & 0.935 & 0.249 \\
  \bottomrule
  \end{tabular}}
  \caption{PID experiment results for varying sampling intervals (left) and confidence thresholds (right).}
  \label{tbl:particle-identification}
\end{table}

In order to achieve further accuracy improvement, measures were taken to minimize the number of misclassified samples by permitting the classifier to withhold decision in cases where errors are likely to occur. This is advantageous especially in applications where the effects of misclassification are less desirable than those of no classification at all. In implementation, the probability of a correct classification is commonly reflected by a confidence measure, which is thresholded at a constant level. For $k$-NN classifiers, this confidence is conservatively defined as $k^j / k$ where $k=7$ is the number of nearest neighbors and $k^j$ denotes the number of training samples from the class $y^j$ among the nearest neighbors in the scope of the current classification instance.

The effects of specific threshold values were examined by performing cross-validation for multiple constant values $c\in\{0.5,0.75,0.9,0.95\}$. For all instances, the best-performing feature model with $n_\text{int}=16$ was selected. In addition to accuracy, the rejection ratio (abbr. RR) was tracked to illustrate the overall confidence in classified samples at various thresholds. Lastly, the calculation of accuracy was constrained to accepted classifications only. Results of the experiment are shown in Table~\ref{tbl:particle-identification} and Figure~\ref{fig:confusion-matrices}.

\section{Conclusions}%
In this work, new analysis methods of Timepix frames were introduced and evaluated. The proposed algorithms update and extend older morphological methods with modern approaches, utilizing ToT information for better accuracy and increased performance. Based on commonly used computer vision methods, the presented algorithms are trans-dimensional and inherently robust to input noise.

In artificial Timepix frames produced from manually annotated experimental datasets, the presented algorithms proved to be viable for linear trajectory fitting, overlap separation and particle identification. On average, SA-RANSAC correctly identified the particle entry and exit points within 3~pixels (165~$\mu\mathrm{m}$) and the azimuthal direction within 2.9~degrees. The modified Hough algorithm achieved predominantly correct results in frames with up to 20 overlaps. The PID feature model was accurate up to 0.893 without rejection and up to 0.935 with rejection rate below 0.25 at confidence threshold 0.9.

\begin{figure}[!htb]
  \centering
  \subfloat[3D-reconstructed pion track observed by a Timepix3 detector, robustly fitted with a linear model (blue) using a generalized RANSAC method~\cite{bergmann20173d,bergmann20193d}.]{\includegraphics[width=0.42\linewidth]{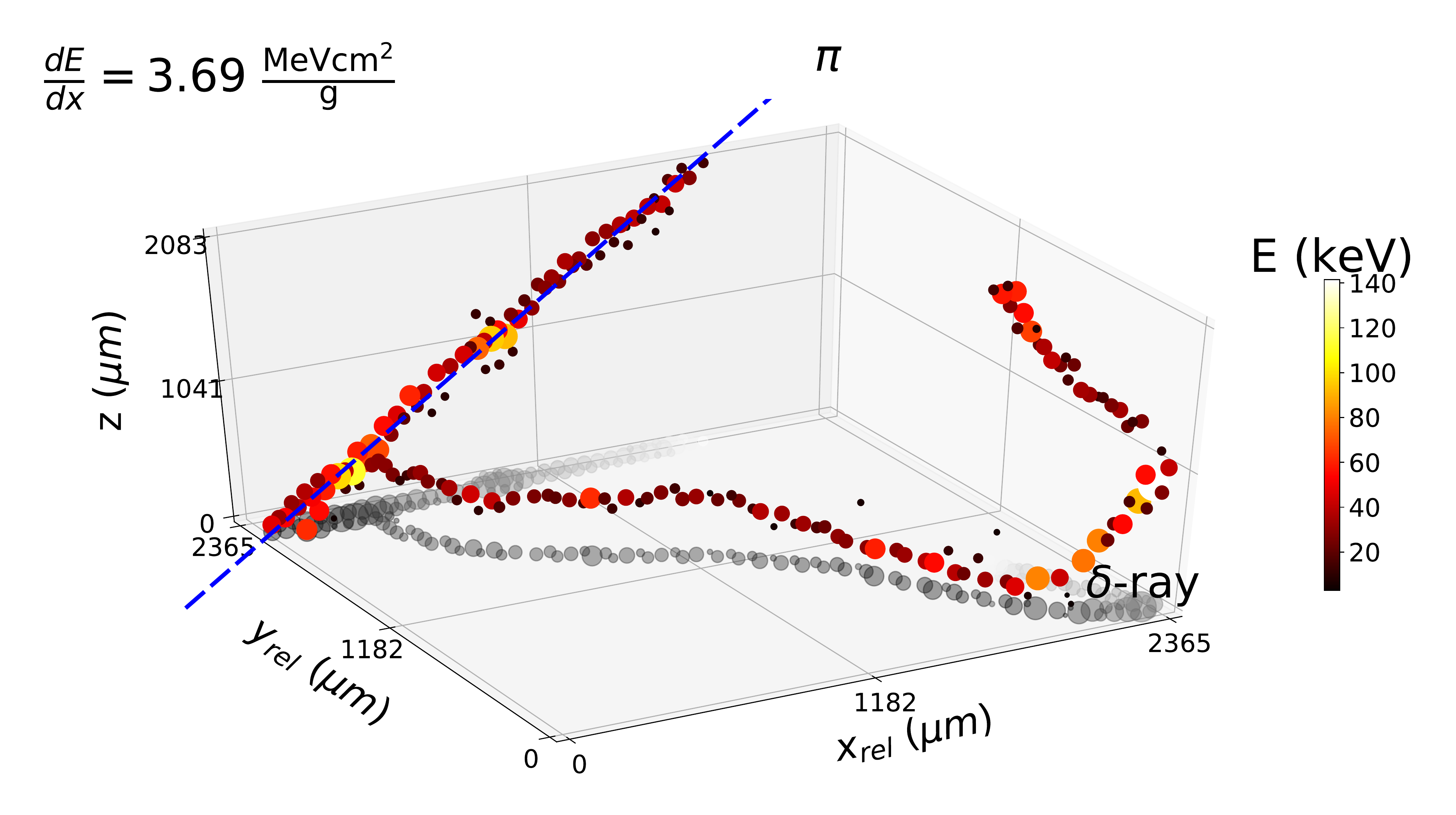}}
  \quad
  \quad
  \subfloat[Histogram of spatial directions fitted from 869,311~tracks observed on July 29-30, 2017 (LHC fill \#6024) by a Timepix detector at MoEDAL Experiment, LHCb~\cite{Manek2018_CTU}.]{\includegraphics[width=0.47\linewidth]{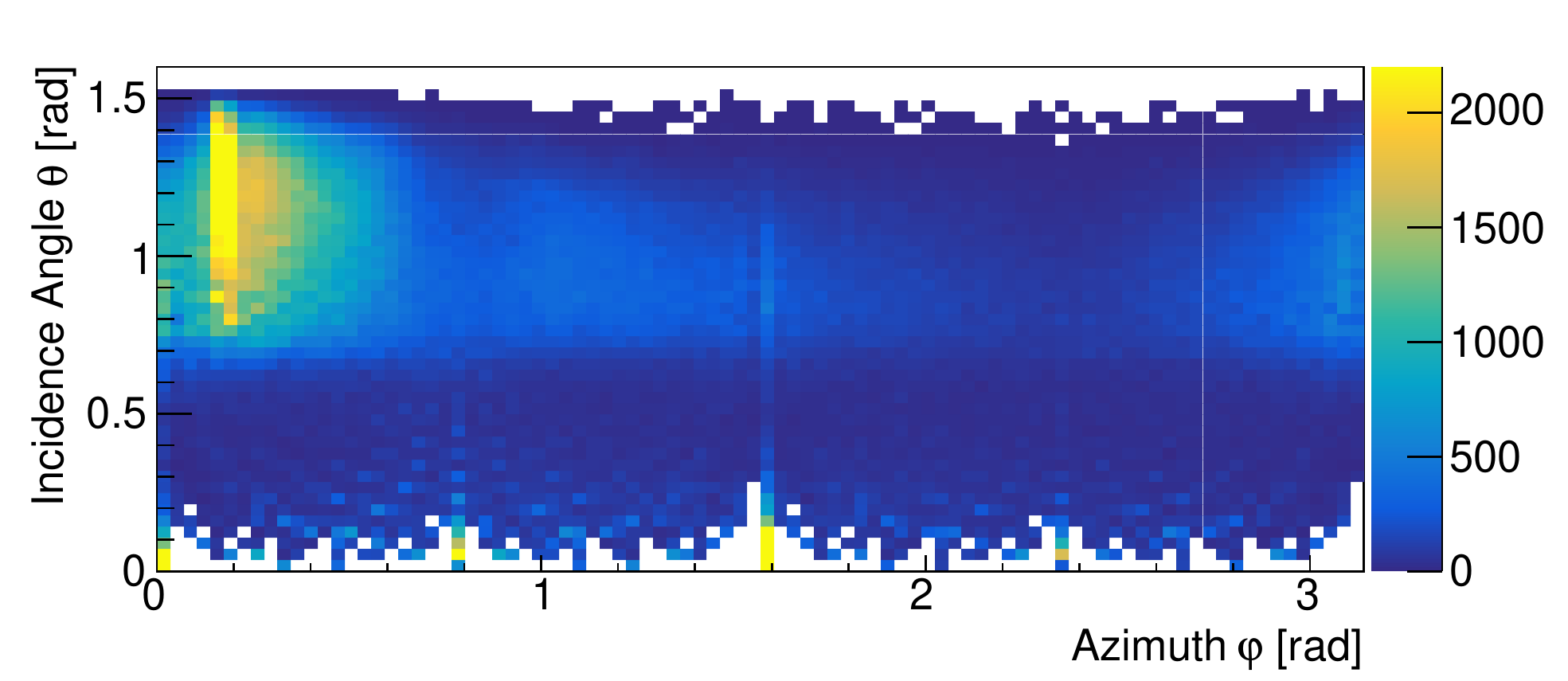}}
  \caption{Selected applications of the presented analysis methods.}
  \label{fig:3d-reconstructed}
\end{figure}

The presented algorithms for segmentation, trajectory reconstruction and particle identification are considered to be viable alternatives to conventional morphological methods, especially in high-flux environments where frame sparsity cannot be necessarily attained. Easily generalizing to 3D, the presented algorithms may find numerous applications in analysis related to Timepix and Timepix3 detectors (illustrated in Figure~\ref{fig:3d-reconstructed}).

\Acknowledgements%
{\small The authors would like to express their sincere gratitude to the Medipix2 and Medipix3 Collaboration for their permanent support. This work was supported by European Regional Development Funds: \textit{"Van de Graaf Accelerator and Tunable Source of Monoenergetic Neutrons and Light Ions"} (No. CZ.02.1.02 / 0.0 / 0.0 / 16\_013 / 0001785) and \textit{"Research Infrastructure for Experiments at CERN"} (LM 2015058)}


{
  \footnotesize
  \bibliographystyle{abbrv}
  \bibliography{eprint}
}


\end{document}